# Interplay of the exciton and electron-hole plasma recombination on the photoluminescence dynamics in bulk GaAs

A. Amo[*], M. D. Martín, L. Viña

*Dept. Física de Materiales, Universidad Autónoma de Madrid, E-28049 Madrid, Spain*

A. I. Toropov, K. S. Zhuravlev

*Institute of Semiconductor Physics, Pr. Lavrentieva, 13, 630090 Novosibirsk, Russia*

We present a systematic study of the exciton/electron-hole plasma photoluminescence dynamics in bulk GaAs for various lattice temperatures and excitation densities. The competition between the exciton and electron-hole pair recombination dominates the onset of the luminescence. We show that the metal-to-insulator transition, induced by temperature and/or excitation density, can be directly monitored by the carrier dynamics and the time-resolved spectral characteristics of the light emission. The dependence on carrier density of the photoluminescence rise time is strongly modified around a lattice temperature of 49 K, corresponding to the exciton binding energy (4.2 meV). In a similar way, the rise-time dependence on lattice temperature undergoes a relatively abrupt change at an excitation density of $120-180 \times 10^{15}$ cm$^{-3}$, which is about five times greater than the calculated Mott density in GaAs taking into account many body corrections.

## I. INTRODUCTION

The carrier dynamics of III-V semiconductors has been extensively studied by means of optical techniques in the past decades.[1] Time-resolved photoluminescence and pump-probe experiments enable the direct observation of the relaxation of carriers photoexcited at different energies and densities above the gap. Both in quantum wells (QWs) and in bulk, most of these studies can be classified in two groups attending to excitation density and lattice temperature conditions: (i) those devoted to the excitonic regime (low lattice temperature and low excitation density -up to $10^{11}$ cm$^{-2}$ in GaAs QWs, and $10^{15}$ cm$^{-3}$ in bulk GaAs-); (ii) those dealing with the electron-hole plasma regime (lattice temperature above the exciton binding energy and/or high excitation densities). However, the photoluminescence (PL) dynamics in the intermediate range, where a Mott transition[2] between the excitonic regime and the conducting electron-hole plasma phase should take place, has not been investigated in detail in III-V semiconductors: a low-resolution time-resolved experiment performed in the early 80's in bulk GaAs (Ref. 3) and a very recent study in 2D (Ref. 4) show that such transition, as the density of photogenerated carriers is increased, is not abrupt. Also recently, time-resolved broadband THz spectroscopy studies have addressed the issue of the exciton to electron-hole plasma transition in QWs by means of intraband differential absorption.[5, 6]

The carrier relaxation dynamics after a pulsed non-resonant excitation is pretty well understood in the aforementioned low and high excitation density regimes. Let us start with the excitonic regime. In QWs, after photocreation of heavy-hole electron pairs, the exciton formation[7, 8] and its relaxation to the bottom of the band result in PL time evolutions with rise times, $t_r$, up to several hundreds of ps long. The excitation density dependence of $t_r$ is strongly influenced by the sample characteristics and the specific excitation conditions of each experiment.[9] Thus, the literature provides a wide spectrum of experimental data with rise times increasing[10] or decreasing[7, 8, 11-13] when raising the excitation density. On the other hand, in bulk III-V samples, these time-resolved studies are scarce[14-17] and ascertain that the free-exciton PL rise time is strongly influenced by trapping in localization centers.

Switching now to the electron-hole plasma regime, both in bulk and QWs, time-resolved studies have concentrated in the thermalization[18-20] and cooling[21, 22] mechanisms of the hot photocreated carriers, but little attention has been paid to the processes responsible for the onset of the luminescence, characterized by its rise time.

The interplay between excitons and free carriers in III-V semiconductors, and their relative contribution to the PL emission at the free-exciton energy, has recently been the subject of intense debate. Time-resolved studies have provided a deep insight on this subject.[23-26] Using a quantum theory of the interaction between photons and an electron-hole population in GaAs

QWs, Kira et al. showed that a Coulomb-correlated unbound electron-hole plasma could reproduce the PL features traditionally assigned to exciton recombination.[23] Recent experiments and their interpretation[24-26] have led to the idea that, in QWs at low temperatures and low/medium excitation densities, excitons constitute a low percentage of the total number of excitations in the system; however, due to the large radiative recombination rate of excitons as compared to that of band-to-band transitions, the exciton emission dominates the PL spectra. Even for densities above the Mott transition, numerical calculations in 1D and 2D systems have shown that both free carriers and excitons coexist and contribute to the PL.[27] Indeed, the competition between the exciton and electron-hole pair contributions to the PL in direct gap semiconductors is still an open question, where time-resolved studies can help to clarify the situation.

In this paper we present a systematic study of the exciton/electron-hole plasma PL dynamics in bulk GaAs in a wide range of lattice temperatures and excitation densities after a pulsed non-resonant excitation. We will concentrate on the onset of the luminescence (rise time) and on the effect of the coexistence of free carriers and excitons on the temporal evolution of the PL. The *excitation-power dependence of the rise time* for different lattice temperatures presents a behavior typical of a metal-to-insulator transition, qualitatively similar to those observed in resistivity measurements in doped bulk semiconductors,[28, 29] in high mobility two-dimensional electron systems,[30-32] or in superconducting thin films.[33-35] This transition, which is continuous but abrupt, takes place at a critical lattice temperature $T_c$. In a similar way, the *lattice-temperature dependence of the rise time* as the excitation density is increased, also undergoes a relatively abrupt change at a critical density $n_c$. Thus, monitoring the PL rise time, we observe a transition that takes place at a density that is about five times greater than the theoretically predicted Mott transition density in photoexcited semiconductors.[36]

## II. EXPERIMENTAL DETAILS

The investigated samples, grown by Molecular Beam Epitaxy, were nominally undoped 2.5 μm GaAs epilayers, encapsulated between two thin AlAs layers to reduce the effects of interface recombination.[37] The samples were mounted on a cold finger cryostat, which enabled a precise control of the lattice temperature in the range 5-100 K, and were non-resonantly photoexcited (1.631 eV) with a Ti:Al$_2$O$_3$ laser that produced 2 ps long pulses. The laser was focused on the sample in a 100 μm-diameter spot. The PL was energy- and time-resolved with a synchroscan streak-camera coupled to a spectrometer. In these experiments, the time and energy resolution of the overall setup is better than 15 ps and 0.3 meV, respectively.

The excitation density of the photogenerated carriers has been calculated considering the measured spot diameter (100±10 μm), the reflectivity and the absorbance of the sample at the energy of the excitation pulses, and the number of photons per laser pulse. Considering all the uncertainties in these quantities we estimate that the carrier densities are correct within a factor of 2. However, the relative uncertainty in the density when comparing two excitation densities within our experiments (just given by the measurement of the laser power) is below 2%.

### III. RESULTS AND DISCUSSION

Figure 1 shows PL spectra recorded 1.8 ns after the excitation at different lattice temperatures, $T_L$, for a low excitation density of 0.75x10$^{15}$ carriers/cm$^3$. For such a long delay, thermodynamical quasiequilibrium between free carriers, excitons and the lattice has been reached.[38] At a lattice temperature of 5 K the spectrum displays the characteristic excitonic emission (1.512-1.516 eV range) and electron-acceptor recombination structures at lower energies, which have been discussed in detail in Ref. 17. For temperatures up to a critical temperature, $T_c$ = 49 K, the spectra are dominated by the excitonic emission. As $T_L$ is increased in the range 5 K ≤ $T_L$ < 49 K the emission from electron-hole pairs becomes apparent (shaded regions) at the band gap energy (indicated by arrows), and its relative intensity increases. In this temperature range, the full width at half maximum (FWHM) of the PL band also increases with temperature (a factor of 2.2 from 5 to 45 K), as it is shown in Fig. 2(b).

For $T_L$ ≥ 49 K the spectra present a much wider overall line-width. Although these temperatures imply energies that are above the exciton binding energy, Coulomb-correlation effects are responsible for the appearance of a wide PL peak at energies below the bandgap.[39] It is remarkable that the spectrum corresponding to $T_L$ = 49 K undergoes an abrupt shift towards lower energies, as can be seen in Fig 2(a), and it is significantly much wider than that at $T_L$ = 45 K. This broadening is observed not only in the overall emission band, but it also becomes

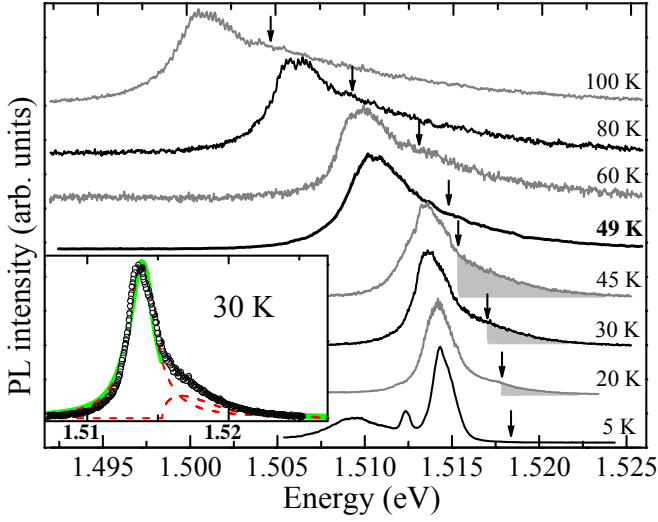
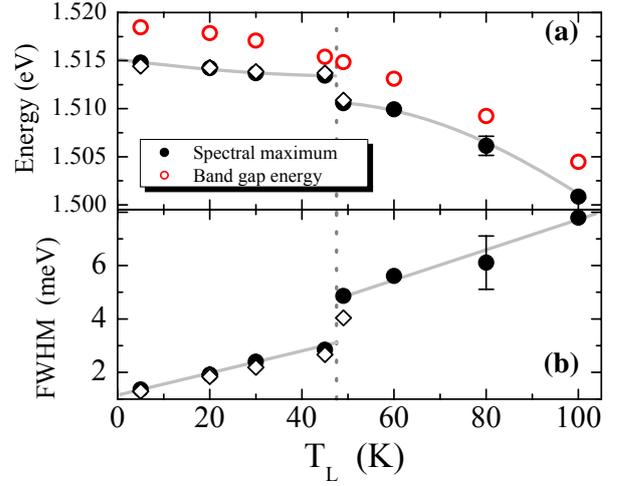

FIG. 1. (Color online) PL spectra recorded 1.8 ns after an initial pulsed excitation density of $0.75\times10^{15}$ cm$^{-3}$ for different lattice temperatures, specified on the side. Arrows indicate the energy position of the band gap at each temperature using the parameters of M. El Allali et al., Phys. Rev. B **48**, 4398 (1993). The shadowed regions show the electron-hole pair luminescence. The inset depicts the 30 K spectrum (open symbols) and the fit to a lorentzian plus band-to-band recombination (green solid line) as described in the text; the dashed lines show these two contributions.

FIG. 2. (Color online) (a) Energy of the spectral maximum extracted from the curves of Fig. 1 (solid points) and that obtained from the lorentzian contribution of the fitted spectra (open diamonds), and energy position of the band gap for each lattice temperatures (open red circles; obtained as in Fig. 1). (b) Full width at half maximum (FWHM) of the whole luminescence band (solid points) and of the lorentzian contribution (open diamonds) for each temperature; the lines are a guide to the eye.

apparent when we focus on each of the contributions to the PL that appear in the spectra shown in Fig. 1. For this purpose a fit is performed in which we use a lorentzian for the excitonic contribution plus a band-to-band recombination (with an appropriate joint density of states and Fermi distributions for electrons and holes). The fits are only meaningful for $T_L$ up to 49 K as for higher temperatures the low energy Coulomb-correlated plasma requires a many-body treatment and cannot be described by a simple lorentzian line-shape. The inset in Fig. 1 depicts the PL (open symbols) together with the fit (solid line) and the two contributions (red dashed lines). Let us note that the energy position and FWHM of the lorentzian contribution (excitonic emission; open diamonds in Fig. 2) present the same features as the overall PL (solid points in

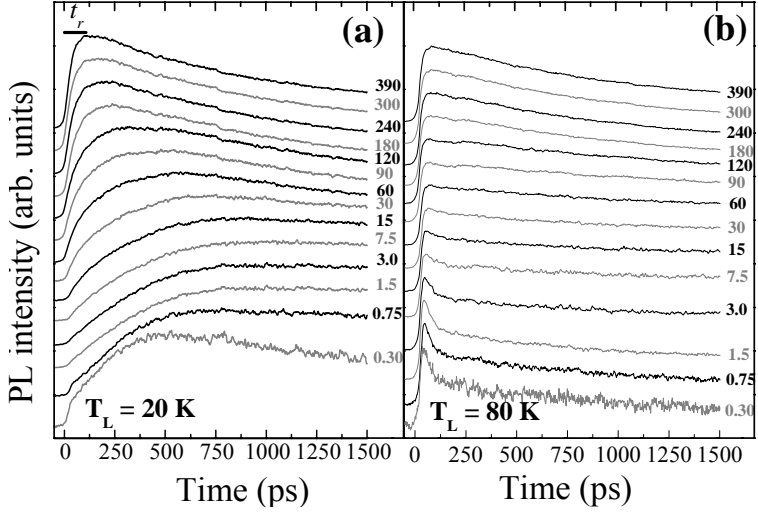

FIG. 3. Time evolution traces at the spectral maximum for $T_L = 20$ K (a) and $T_L = 80$ K (b). The numbers on the right side of each panel show the excitation density for each trace in units of $10^{15}$ cm$^{-3}$.

Fig. 2), including the abrupt broadening at $T_L = 49$ K [a factor of 1.5 (1.7) from $T_L = 45$ K to $T_L = 49$ K in the excitonic (overall) PL band].

The abrupt shift and the broadening at $T_c$ demonstrate that there are two kinds of spectra belonging to two different regimes, as we shall discuss below. Moreover, Fig. 2(a) shows that the exciton emission energy approaches the band gap at $T_c$, indicating the vanishing of the exciton binding energy, i.e. the exciton ionization, at this lattice temperature.

Figure 3 depicts PL time-evolution traces at the energy of the spectral maximum for low [20 K; (a)] and high [80 K; (b)] $T_L$ at different excitation densities $n$. A proper rate equation model would be necessary to describe this time evolution. However, the modeling of the electron-hole plasma contribution to the luminescence at the spectral maximum is very complicated.[23, 24] Thus, following the usual practice in the literature, we employ the time for the PL to reach its maximum intensity, defined as $t_r$, to analyze the initial emission dynamics. $t_r$ is shown for the uppermost curve in Fig. 3(a) with a horizontal bar. Let us start discussing the high excitation-density regime. The temporal traces at these densities ($n > 150 \times 10^{15}$ cm$^{-3}$) are qualitatively very similar for both lattice temperatures: for such high densities the system behaves like an electron-hole plasma due to the effective carrier screening, and the initial carrier temperatures are much larger than $T_L$. A detailed analysis of the traces shows that for $T_L = 80$ K the rise times are shorter due to the enhancement of the phonon-assisted relaxation of carriers in the bands, and to the opening of non-radiative recombination channels associated to carrier-phonon scattering at high lattice temperatures. On the other hand, at low excitation densities the PL time evolution presents very different features at $T_L = 20$ K and $T_L = 80$ K, as easily seen in Fig. 3.

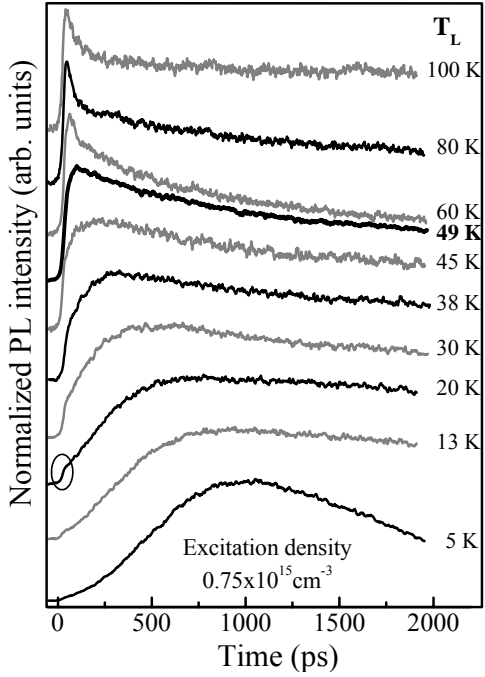

FIG. 4. Time evolution traces at the spectral maximum of the luminescence for different lattice temperatures ($T_L$) at an excitation density of $0.75 \times 10^{15}$ cm$^{-3}$. The circle encloses the *fast component* of the onset of the PL for $T_L = 20$ K.

The shape and characteristic times of the onset of the PL, at low excitation densities, are strongly dependent on the lattice temperature. Figure 4 depicts temporal traces for an excitation of $0.75 \times 10^{15}$ cm$^{-3}$ at different $T_L$ at the spectral maximum. For temperatures up to 45 K the onset of the luminescence is characterized by two distinct features: (i) a *fast* initial *component* (enclosed by a circle in the trace of 20 K); and (ii) a subsequent *slower rise*. The interplay between the exciton and electron-hole pair emissions is responsible for the shape of the time evolution of the onset of the PL at the free-exciton energy.

This interplay results in a competition between the *fast component* (related to electron-hole pair recombination) and the *slow component* (excitonic recombination) in the onset of the luminescence, as we will discuss in the following paragraphs.

At low excitation densities, such as the one corresponding to Fig. 4, the *slow component* fully dominates the onset of the PL at the lowest $T_L$ (5 K). Under these conditions the PL is mostly arising from excitonic recombination.[9, 24] The long $t_r$ reflects the slow phonon-assisted exciton relaxation from states of large momentum $k$, where electron and holes were bound to form excitons, to the radiatively active states at $k = 0$.[14] The *fast component*, already seen at 13 K, has been previously observed in GaAs and tentatively attributed either to the emission of free electron-hole pairs[15] or to a rapid exciton formation mediated by LO-phonon interactions.[40] Our results discard the latter mechanism since the fast component is absent at the lowest temperature and LO-phonon emission is temperature independent. Therefore, the *fast component* in the rise can be unambiguously attributed to the recombination of unbound electron-hole pairs.

As $T_L$ is increased, the *fast component* becomes more important, as can be seen in Fig. 4. Therefore, the fraction of excitons present in the system at short times is reduced when $T_L$ is increased, which is in agreement with the computational results of Koch et *al*. presented in Ref. 27. For $T_L > T_c = 49$ K, the *fast initial component* fully dominates the rise time. For those values of $T_L$ excitons are ionized (as $k_B T_L > 4.2$ meV, the exciton binding energy, for $T_L > 49$ K) and all the luminescence arises from the recombination of Coulomb-correlated electron-hole pairs (see Fig. 1).

We have just discussed the behavior of the *fast rising component* of the PL for a given low excitation density as a function of $T_L$. Let us now examine this dependence at a given $T_L$ for different excitation powers. For a low $T_L$ [i.e., 20 K; Fig. 3(a)], where exciton formation is not inhibited by thermal ionization, the fraction of electron-hole pairs that bind to form excitons increases with increasing excitation density.[9, 41] As a result, the *fast component* in the PL rise (electron-hole recombination) is overcome by the *slow* excitonic *component* when the excitation density is increased ($0.3 \times 10^{15} < n < 3.0 \times 10^{15}$ cm$^{-3}$ at short times), as borne out by our experiments. However, with a further increase in the excitation density, screening between carriers starts to be an important factor and inhibits the binding of electron-hole pairs into excitons;[27] electron-hole pair recombination is again important and the dynamics accelerate. At the highest densities ($n > 150 \times 10^{15}$ cm$^{-3}$) the emission occurs mainly from electron-hole pair recombination.

The preceding discussion shows that the interplay between the exciton recombination and electron-hole pair emission determines the shape of the time evolution of the onset of the PL at the free-exciton energy. At the shortest times the emission comes mainly from electron-hole pair recombination (*fast component*). Its relative contribution to the PL, compared to the excitonic one (*slow component*), increases when $T_L$ is increased and decreases when the excitation density is increased, as long as the excitation density is kept below $15 \times 10^{15}$ cm$^{-3}$. With a further increase in excitation density (above $150 \times 10^{15}$ cm$^{-3}$) the system is populated by electron-hole pairs as exciton formation is hindered by screening.

Figure 4 also shows a striking feature in the high $T_L$ temporal evolutions. For $T_L$ above 49 K the fast component of the onset of the PL is followed by an initial fast decay, which is more evident as $T_L$ is increased. If we focus on the $T_L = 80$ K case, Fig. 3(b) reveals that this fast initial drop is more important at low excitation densities, being completely absent for $n > 50 \times 10^{15}$ cm$^{-3}$. The origin of this initial fast decay may be sought in the warming of the electron-hole plasma. At the lowest excitation densities, the fast subpicosecond thermalization[18, 19] in conjunction with efficient LO-phonon assisted relaxation, results in thermalized carrier populations with initial temperatures close but slightly below $T_L$. Figure 5 shows the initial carrier temperature $\langle T \rangle$ (averaged over the first 25 ps, i. e., just before the initial fast decay) as a function of excitation density for $T_L = 80$ K. The temperature was extracted from the high energy tail of the PL assuming, for the sake of simplicity, Boltzmann distributions.[42] Indeed, for excitation densities below $50 \times 10^{15}$ cm$^{-3}$ the initial carrier temperature is lower than $T_L$. We interpret the initial fast decay of the PL as a consequence of the warming of the carriers to $T_L$, which changes the carrier distributions, in particular resulting in a depletion of the states at the energy of the maximum of the PL band. This effect has been observed for excitons in GaAs QWs.[13] For $n > 50 \times 10^{15}$ cm$^{-3}$ this initial drop is absent since the carrier temperature is above $T_L$ and, therefore, the depletion does not occur.

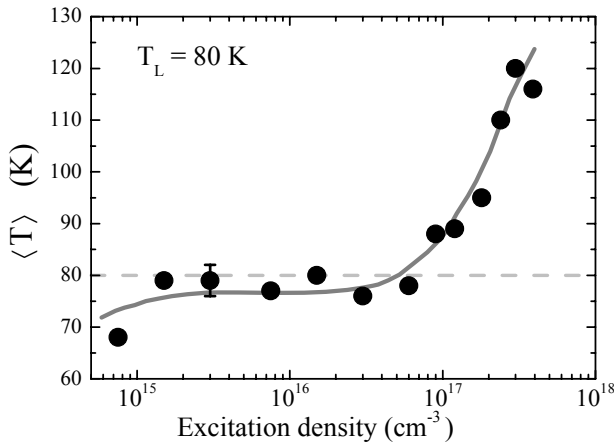

FIG. 5. Initial carrier temperature, $\langle T \rangle$ (averaged over the first 25 ps), as a function of excitation density for $T_L = 80$ K; the solid line is a guide to the eye.

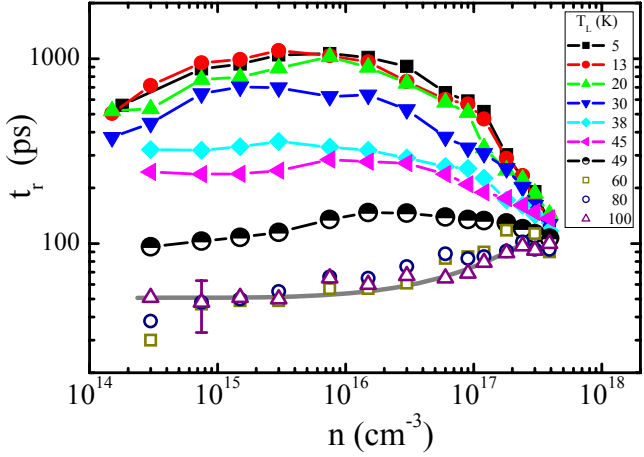 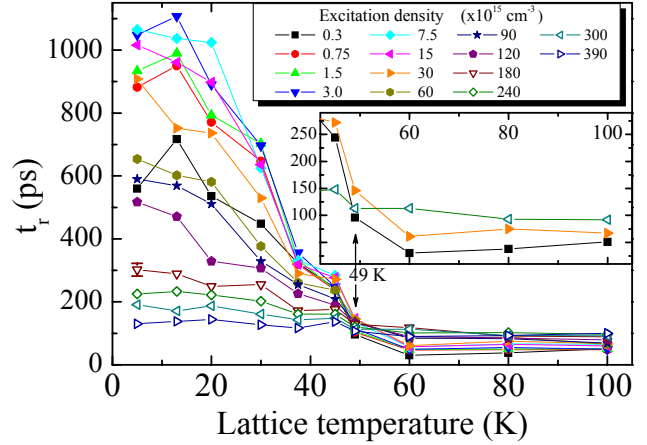

FIG. 6. (Color online) Rise time as a function of excitation density for different lattice temperatures. Solid symbols correspond to the excitonic/insulating phase; open symbols correspond to the electron-hole/metallic phase. The solid grey line is a guide to the eye.

FIG. 7. (Color online) Rise time versus lattice temperature for each set of excitation densities (note that excitation densities are in units of $10^{15}$ cm$^{-3}$). The inset shows the temperature range 42 K - 105 K in greater detail for excitation densities $0.3 \times 10^{15}$ cm$^{-3}$ (squares), $30 \times 10^{15}$ cm$^{-3}$ (solid triangles) and $300 \times 10^{15}$ cm$^{-3}$ (open triangles).

The main point we can extract from all the above discussions is that a critical temperature $T_c$ = 49 K can be identified, which sets a boundary in the spectral (Fig. 1) and dynamic (Figs. 3 & 4) behavior of carriers in the system. In the case of the dynamics, the interplay between the excitonic (*slow component*) and electron-hole pair (*fast component*) recombination is summarized in Figs. 6 and 7 for different $T_L$ and excitation densities. Figure 6 depicts $t_r$ as a function of excitation density, $n$, for different lattice temperatures. The curves can be classified in two groups, corresponding to $T_L < T_c$ (solid symbols) and $T_L > T_c$ (open symbols), plus the curve at 49 K. The $t_r$ dependence on $n$ of these groups shows certain symmetry with respect to the $T_L = T_c$ curve, with negative curvature for $T_L < T_c$ and positive for $T_L > T_c$. This symmetric behavior is characteristic of a metal-to-insulator transition,[28, 31] with $T_L$ as the order parameter having a critical value of 49 K (in resistivity measurements in doped semiconductors the order parameter in the metal-to-insulator transition is the electron density). The lattice temperature that sets the boundary between the insulator and metal behavior in $t_r$ ($T_c$ = 49 K) coincides with the exciton binding energy (i. e., $k_B T_c$ = 4.2 meV). This fact evidences that ionization is dominated by exciton-phonon interaction rather than by exciton-exciton scattering.

The first group of curves in Fig. 6, $T_L < 49$ K (solid symbols), corresponds to the excitonic insulating phase in which the onset of the PL is dominated by the *slow component* and $t_r$ is governed by the exciton relaxation. For $n < 3 \times 10^{15}$ cm$^{-3}$ and $T_L$ up to 30 K, the monotonic increase of $t_r$ with density is related to the trapping of free excitons in bound states.[17] For higher excitation densities, the steady decrease of $t_r$ with increasing density arises from the fast relaxation of excitons induced by exciton-exciton scattering.[7] The second group, $T_L > 49$ K (open symbols), corresponds to the metallic phase, with the onset of the PL dominated by the *fast component* (electron-hole pair recombination). The rise time, which increases monotonically with increasing $n$, is determined by the thermalization and cooling of carriers. This increase can be qualitatively explained taking into account the initial carrier temperature. For densities above ~$50 \times 10^{15}$ cm$^{-3}$, the temperature of the initial electron-hole plasma is higher than $T_L$ (see Fig. 5); carriers cool down to $T_L$ through carrier-phonon interaction and it takes some time to reach the highest occupation of the lowest energy states, resulting in a delay for the PL to reach its maximum.[13] As the excitation density is raised, the initial carrier temperature is higher and the cooling takes longer, leading to an increase of $t_r$.[22, 43, 44]

Only at the highest excitation densities, in the region where all the curves tend to approach a common value of $t_r \approx 100$ ps, the rise time is essentially characterized by electron-hole recombination for any lattice temperature, due to the effective carrier screening as already discussed above. In order to reinforce this idea we have plotted, in Fig. 7, the rise-time dependence on $T_L$ for several excitation densities. There is a temperature (49 K, the critical temperature) for which $t_r$ is nearly independent of the excitation density. Moreover, the curves reverse their order when crossing this temperature (see inset of Fig. 7; only 3 curves are shown for clarity). These two facts are qualitatively identical to what is found in resistivity studies around the metal-to-insulator transition in two dimensional high mobility semiconductors.[31, 32]

Figure 7, shows again two groups of curves, separated by an excitation density of ~$150 \times 10^{15}$. For $n \leq 120 \times 10^{15}$ cm$^{-3}$ (solid points) and $T_L < 49$ K, the dependence of the rise time on lattice temperature reflects the aforementioned interplay between exciton and electron-hole pair recombination. For $n \geq 180 \times 10^{15}$ cm$^{-3}$ (open points) the rise-time dependence on $T_L$ shows a behavior much less dependent on excitation density, as in this regime the carriers in the system form an electron-hole plasma (metallic state). Thus, the excitation density range 120-180$\times 10^{15}$

cm$^{-3}$ establishes a phase boundary in the characteristics of the onset of the PL, similar to the $T_L$ = 49 K boundary discussed above. The metal-to-insulator transition in the system is set by this density range (densities varying by a factor 1.5), much more abrupt than the observed Mott transition in recent experiments in QWs,[4] which takes place over an order of magnitude in excitation densities. The transition densities we find are about five times greater than the theoretical calculations for the Mott transition by Haug and Schmitt-Rink.[36] Their calculations, making use of Hartree-Fock and self-screening corrections to the exciton energy, yield a Mott density for GaAs of 28.1x10$^{15}$ cm$^{-3}$ at $T_L$ = 0.

Finally, let us further comment on the relation between the exciton binding energy and the density and temperatures at which the Mott transition takes place. This can be readily investigated in QWs, where the binding energy can be controlled by varying the well width. One could argue that higher binding energies would lead to higher critical lattice temperatures for the metal-to-insulator transition. Theoretical calculations[45] show that for GaAs/AlGaAs QWS, both the binding energy and the critical transition density increase when the well width is reduced, confirming that a stronger exciton is harder to dissociate. However, the scenario is rather complex as demonstrated by calculations showing a transition temperature varying from 40 K to 80 K when the carrier density is increased in a given GaAs/AlGaAs QW.[46]

## IV. SUMMARY

In conclusion, we have demonstrated that both the spectral and carrier dynamics properties of bulk GaAs at various lattice temperatures and excitation densities are dominated by the interplay between exciton and electron-hole pair recombination and relaxation. Both contributions to the PL coexist and cannot be separated at low temperatures (below 49 K) and low excitation densities (below 120-180x10$^{15}$ cm$^{-3}$). We have shown evidence of a continuous but rather abrupt metal-to-insulator transition in the rise-time characteristics, at a lattice temperature of 49 K. This temperature is very close to that associated to the exciton binding energy in bulk GaAs (48.7 K, 4.2 meV respectively). This fact suggests that the transition is ruled by exciton-phonon interaction rather than by exciton-exciton scattering. Similarly, an excitation density of 120-180x10$^{15}$ cm$^{-3}$ sets a transition in the rise-time dependence on lattice temperature, which is about five times greater than the theoretical Mott transition density.

## ACKNOWLEDGEMENTS

We thank C. Tejedor for fruitful discussion. This work was partially supported by the Spanish MCYT (MAT2002-00139), the Comunidad Autónoma de Madrid (GR/MAT/0099/2004) and the Russian Basic Research Fund (grant no. 04-02-16774a). A. A. acknowledges a scholarship of Spanish Secretaría de Estado de Educación y Universidades (MEC).

## REFERENCES AND FOOTNOTES

[*] Author to whom correspondence should be addressed; electronic mail: alberto.amo@uam.es

[1] J. Shah, Ultrafast spectroscopy of semiconductors and semiconductor nanostructures (Springer-Verlag, Berlin, Germany, 1996)

[2] N. F. Mott, Proc. Phys. Soc. A **62**, 416 (1949).

[3] E. O. Göbel, P. H. Liang and D. v. d. Linde, Solid State Commun. **37**, 609 (1981).

[4] L. Kappei, J. Szczytko, F. Morier-Genoud and B. Deveaud, Phys. Rev. Lett. **94**, 147403 (2005).

[5] R. A. Kaindl, M. A. Carnahan, D. Hägele, R. Lövenich and D. S. Chemla, Nature **423**, 734 (2003).

[6] R. Huber, R. A. Kaindl, B. A. Schmid and D. S. Chemla, cond-mat/0508082 (2005).

[7] T. C. Damen, J. Shah, D. Y. Oberli, D. S. Chemla, J. E. Cunningham and J. M. Kuo, Phys. Rev. B **42**, 7434 (1990).

[8] R. Kumar, A. S. Vengurlekar, S. S. Prabhu, J. Shah and L. N. Pfeiffer, Phys. Rev. B **54**, 4891 (1996).

[9] J. Szczytko, L. Kappei, J. Berney, F. Morier-Genoud, M. T. Portella-Oberli and B. Deveaud, Phys. Rev. Lett. **93**, 137401 (2004).

[10] H. W. Yoon, D. R. Wake and J. P. Wolfe, Phys. Rev. B **54**, 2763 (1996).

[11] L. Muñoz, E. Pérez, L. Viña and K. Ploog, Phys. Rev. B **51**, 4247 (1995).

[12] P. Roussignol, C. Delalande, A. Vinattieri, L. Carraresi and M. Colocci, Phys. Rev. B **45**, R6965 (1992).

[13] R. Eccleston, R. Strobel, W. W. Rühle, J. Kuhl, B. F. Feuerbacher and K. Ploog, Phys. Rev. B **44**, R1395 (1991).

[14] M. Gurioli, P. Borri, M. Colocci, M. Gulia, F. Rossi, E. Molinari, P. E. Selbmann and P. Lugli, Phys. Rev. B **58**, R13403 (1998).

[15] R. Höger, E. O. Göbel, J. Kuhl, K. Ploog and H. J. Quiesser, J. Phys. C **17**, L905 (1984).

[16] J. X. Shen, R. Pittini, Y. Oka and E. Kurtz, Phys. Rev. B **61**, 2765 (2000).

[17] A. Amo, M. D. Martin, L. Klopotowski, L. Viña, A. I. Toropov and K. S. Zhuravlev, Appl. Phys. Lett. **86**, 111906 (2005).

[18] L. Rota, P. Lugli, T. Elsaesser and J. Shah, Phys. Rev. B **47**, 4226 (1993).

[19] A. Alexandrou, V. Berger and D. Hulin, Phys. Rev. B **52**, 4654 (1995).

[20] W. H. Knox, in Hot Carriers in Semiconductor Nanostructures: Physics and Applications, edited by J. Shah (Academic Press, Inc., Boston, San Diego, New York, London, Sydney, Tokio, Toronto, 1992).

[21] D. von der Linde and R. Lambrich, Phys. Rev. Lett. **42**, 1090 (1979).

[22] K. Leo, W. W. Rühle, H. J. Queisser and K. Ploog, Phys. Rev. B **37**, 7121 (1988).

[23] M. Kira, F. Jahnke and S. W. Koch, Phys. Rev. Lett. **81**, 3263 (1998).

[24] S. Chatterjee, C. Ell, S. Mosor, G. Khitrova, H. M. Gibbs, W. Hoyer, M. Kira, S. W. Koch, J. P. Prineas and H. Stolz, Phys. Rev. Lett. **92**, 067402 (2004).

[25] J. Szczytko, L. Kappei, J. Berney, F. Morier-Genoud, M. T. Portella-Oberli and B. Deveaud, Phys. Rev. B **71**, 195313 (2005).

[26] D. Robart, X. Marie, B. Baylac, T. Amand, M. Brousseau, G. Bacquet, G. Debart, R. Planel and J. M. Gerard, Solid State Commun. **95**, 287 (1995).

[27] S. W. Koch, W. Hoyer, M. Kira and V. S. Filinov, Phys. Stat. Sol. (b) **238**, 404 (2003).

[28] G. A. Thomas, M. Paalanen and T. F. Rosenbaum, Phys. Rev. B **27**, R3897 (1983).

[29] T. F. Rosenbaum, R. F. Milligan, M. A. Paalanen, G. A. Thomas, R. N. Bhatt and W. Lin, Phys. Rev. B **27**, 7509 (1983).

[30] S. V. Kravchenko, G. V. Kravchenko, J. E. Furneaux, V. M. Pudalov and M. D'Iorio, Phys. Rev. B **50**, 8039 (1994).

[31] S. V. Kravchenko, W. E. Mason, G. E. Bowker, J. E. Furneaux, V. M. Pudalov and M. D'Iorio, Phys. Rev. B **51**, 7038 (1995).

[32] H. W. Jiang, C. E. Johnson, K. L. Wang and S. T. Hannahs, Phys. Rev. Lett. **71**, 1439 (1993).

[33] D. B. Haviland, Y. Liu and A. M. Goldman, Phys. Rev. Lett. **62**, 2180 (1989).

[34] S. R. Khan, E. M. Pedersen, B. Kain, A. J. Jordan and R. P. Barber, Phys. Rev. B **61**, 5909 (2000).

[35] J. M. Valles, Jr., Shih-Ying Hsu, R. C. Dynes, J. P. Garno, Physica B **197**, 522 (1994).

[36] H. Haug and S. Schmitt-Rink, Prog. Quant. Electr. **9**, 3 (1984).

[37] K. S. Zhuravlev, A. I. Toropov, T. S. Shamirzaev and A. K. Bakarov, Appl. Phys. Lett. **76**, 1131 (2000).

[38] Thermal equilibrium between carriers and the lattice may not be reached at the lowest temperatures ($T_L < 20$ K) before all carriers have radiatively recombined, as found in QWs (Refs. 9, 22). However, thermodynamic equilibrium between free carriers and excitons is present at all times (see Ref. 26).

[39] E. Grilli, M. Guzzi, R. Zamboni and L. Pavesi, Phys. Rev. B **45**, 1638 (1992).

[40] I. Reimand and J. Aaviksoo, Phys. Rev. B **61**, 16653 (1999).

[41] P. E. Selbmann, M. Gulia, F. Rossi, E. Molinari and P. Lugli, Phys. Rev. B **54**, 4660 (1996).

[42] W. W. Rühle and H.-J. Polland, Phys. Rev. B **36**, 1683 (1987).

[43] P. Langot, R. Tommasi and F. Vallée, Solid State Commun. **98**, 171 (1996).

[44] N. Del Fatti, P. Langot, R. Tommasi and F. Vallée, Phys. Rev. B **59**, 4576 (1999).

[45] E. X. Ping and H. X. Jiang, Phys. Rev. B **47**, 2101 (1993).

[46] S. Ben-Tabou de-Leon and B. Laikhtman, Phys. Rev. B **67**, 235315 (2003).